\documentclass[twocolumn,pra,aps,showpacs]{revtex4}
\usepackage{amssymb}
\usepackage{amsmath}
\usepackage{epsfig}

\begin{document}

\title{Nonlinear Ramsey interferometry with the Rosen-Zener pulses
 on a two-component Bose-Einstein condensate}
\author{ Sheng-Chang Li$^{1,2}$}
\author{ Li-Bin Fu$^{2}$ }
\author{ Wen-Shan Duan$^{1}$ }
\author{ Jie Liu$^{2,3,}$ }
\email[Email: ]{liu_jie@iapcm.ac.cn}

\affiliation{1.Physics and Electronic Engineering College, Northwest
Normal University, 730070, Lanzhou,
 People's Republic of China\\
2.Institute of Applied Physics and Computational Mathematics,
 100088, Beijing, People's Republic of China\\
3.Center for Applied Physics and Technology, Peking University,
100084, Beijing, People's Republic of China}

\begin{abstract}
We propose a feasible scheme to realize  nonlinear Ramsey
interferometry with a two-component Bose-Einstein condensate, where
the nonlinearity arises from the interaction between coherent atoms.
In our scheme, two Rosen-Zener pulses are separated by an
intermediate holding period of variable duration and through varying
the holding period we have observed nice Ramsey interference
patterns in the time domain. In contrast to the standard Ramsey
fringes our nonlinear Ramsey patterns display diversiform structures
ascribed to the interplay of the nonlinearity and asymmetry. In
particular, we find that the frequency of the nonlinear Ramsey
fringes exactly reflects the strength of nonlinearity as well as the
asymmetry of system. Our finding suggests a potential application of
the nonlinear Ramsey interferometry in calibrating the atomic
parameters such as scattering length and energy spectrum.

\end{abstract}

\pacs{37.25.+k,67.85.Fg,03.75.Lm,03.75.-b} \maketitle

\section{Introduction}
The technique of Ramsey interferometry with separated oscillating
fields was first proposed to investigate the molecular beam
resonance \cite{Ram1950}. The key feature of the observed Ramsey
pattern in the frequency domain is that the width of the central
peak is determined by the inverse of the time taken by the particle
to cross the intermediate drift region \cite{PRA76(2007)two-level}.
Indeed, the Ramsey interference experiments can be operated either
in the time domain with temporally separated pulses and fixed
particle or in the space domain with spatially separated fields and
moving particle \cite{PRA75(2007)two-frequency}. The Ramsey's
interferometric method provides
 the basis of atomic fountain clocks that now serve as time standards
  \cite{PRL82(1999),PRL85(2000)} and stimulates the rapid advancement
  in the field of precision
measurements in atomic physics. Since applying the laser cooling
techniques to trapped atoms, the atom interferometers with cold
atoms have been used to measure rotation \cite{rotation1},
gravitational acceleration \cite{rotation2,acceleration}, atomic
fine-structure constant \cite{PRL70(1993)}, atomic recoil frequency
\cite{PRA67(2003)}, and atomic scattering properties
\cite{PRL92(2004)}, to name only a few.

 On the
other hand, the experimental realization of the Bose-Einstein
condensate (BEC) in a dilute atomic gas \cite{BECfind1,BECfind2}
brings a fascinating opportunity for the purpose of precision
measurement due to the very slow atoms and changes the prospects of
frequency standards entirely. Recently, Ramsey fringes between atoms
and molecules in time domain have been observed by using trapped BEC
of $^{85}$Rb atoms \cite{AMRam1} in experiment. This offers the
possibility of precise measurement of binding energy of the
molecular state in BEC \cite{AMRam2,AMRam3}.

With the development of atom interferometry techniques, researchers
are seeking to exploit new interferometric methods using trapped
 BEC \cite{InBECs,chli1}. With the emergence of the nonlinear interaction
 between the coherent ultracold atoms,
 the BECs show marvelous nonlinear tunneling and interference properties that
 are distinguished from the traditional quantum systems.
Motivated by our recent study on nonlinear Rosen-Zener (RZ)
transition \cite{Ye}, in this paper we construct a nonlinear Ramsey
interferometer with applying a sequence of two identical nonlinear
RZ tunneling processes (i.e., RZ pulses). The RZ model was first
proposed to study the spin-flip of two-level atoms interacting with
a rotating magnetic field to explain the double Stern-Gerlach
experiments \cite{RZT}. Differing from the Landau-Zener model
\cite{LZT}, RZ model has set the energy difference between two modes
as a constant whereas the coupling strength is time dependent. In
our interferometry scheme, two RZ pulses are separated by a
intermediate holding period of variable duration and through varying
the holding period we have observed diversiform Ramsey interference
patterns in contrast to the standard Ramsey fringes. Using a simple
nonlinear two-mode model, we thoroughly investigate the physics
underlying the interference patterns both numerically and
analytically.  We find that the frequency of the nonlinear Ramsey
fringes exactly reflects the strength of nonlinearity as well as the
asymmetry of system. This observation suggests an potential
application in calibrating the atom parameters such as scattering
length and energy spectrum via measuring the frequency of Ramsey
fringes.

Our paper is organized as follows. In Sec. II, we present our
nonlinear Ramsey interferometer and demonstrate diversiform
interference patterns. In Sec. III, we make detailed theoretical
analysis on the nonlinear Ramsey interferometry. In the sudden limit
and adiabatic limit, we have derived analytically the frequencies of
the fringes in time domain and their dependence of the atomic
parameters. Sec. IV is our discussions and applications, where we
also extend our discussions to the double-well BEC systems.

\section{Nonlinear Ramsey interferometry}

\subsection{Interferometer scheme}
We consider that a condensate, for example, $^{87}$Rb atoms in a
magnetic trap are driven by a microwave coupling into a linear
superposition of two different hyperfine states, i.e., $F=1,m_F=-1$
and $F=2,m_F=+1$. A near resonant pulsed radiation laser field is
used to couple the two internal states.
The total density and mean phase remain constant during the
condensate evolution. Within the standard rotating-wave
approximation, for any one pulse the Hamiltonian describing the
transition between the two internal states can be read ($\hbar =1$)
\begin{equation}\label{secondH}
\hat{H}=-\frac{\gamma}{2}(\hat{a}^\dagger\hat{a}-\hat{b}^\dagger\hat{b})
-\frac{c}{4}(\hat{a}^\dagger\hat{a}-\hat{b}^\dagger\hat{b})^2
+\frac{v}{2}(\hat{a}^\dagger\hat{b}+\hat{b}^\dagger\hat{a}),
\end{equation}
where $\hat{a}$ ($\hat{b}$) and $\hat{a}^\dagger$
($\hat{b}^\dagger$) are boson annihilation and creation operators
for two components, respectively.
$\gamma=-\delta+(4N\pi\hbar^2/m)(a_{11}-a_{22})\eta$ is the energy
difference between two states characterizing the asymmetry of the
system, $c=(2\pi\hbar^2/m)(a_{11}+a_{22}-2a_{12})\eta$ is the
nonlinear strength describing atomic interactions, and $v$ denotes
the coupling strength which is proportional to the intensity of
near-resonant laser field. $\delta$ is the detuning of lasers from
resonance, $a_{ij}$ is the $s-$wave scattering amplitude of
hyperfine species $i$ and $j$, $\eta$ is a constant of order 1
independent of the hyperfine index, relating to an integral of
equilibrium condensate wave function \cite{RMP73(2001)}, $N$ is the
atom number, and $m$ is the mass of atom.

In the limit of large particle number, the  operators in the above
field equations could be replaced by the complex numbers, we thus
obtain following mean-field equations that  describe the evolution
of the above two-component BEC system  effectively,
\begin{equation}\label{seq}
i {d\over d t}\left(
\begin{array}{c}
a \\b
\end{array}
\right) =H(v)\left(
\begin{array}{c}
a \\b
\end{array}
\right),
\end{equation}
with the Hamiltonian
\begin{equation}\label{H}
H(v)=\left(
\begin{array}{cc}
\frac{\gamma}{2}+\frac{c}{2}(|b|^2-|a|^2) & \frac{v}{2}\\
 \frac{v}{2} & -\frac{\gamma}{2}-\frac{c}{2}(|b|^2-|a|^2)\\
\end{array}
\right),
\end{equation}
where $a$ and $b$ denote the amplitudes of probabilities for two
components and the total probability $|a|^2+|b|^2=1$.

\begin{figure}[h]
\begin{center}
\rotatebox{0}{\resizebox *{8.cm}{3.cm} {\includegraphics
{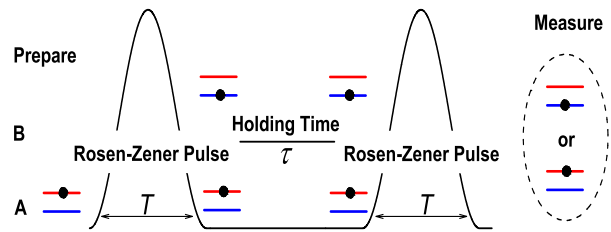}}}
\end{center}
\caption{(Color online) Schematic plot of nonlinear Ramsey
interferometer with two-component trapped BEC in time domain,
starting with a RZ pulse, addition of a holding period, ending with
another RZ pulse.}\label{scheme}
\end{figure}

Using the above two-component BEC system we are capable to realize a
nonlinear Ramsey interferometer, in which the nonlinearity
represents the interparticle interaction. The main structure of our
nonlinear Ramsey interferometer is illustrated by Fig. \ref{scheme},
in which the variation of the coupling strength is governed by two
Rosen-Zener pulses of the form:
\begin{equation}\label{v}
v(t)=\left\{
\begin{array}{ll}
0,& t<0;\\
 v_0\sin^2(\frac{\pi t}{T}), & t\in[0,T];\\
 0, & t\in(T,T+\tau);\\
  v_0\sin^2[\frac{\pi(t-T-\tau)}{T}], & t\in[T+\tau,2T+\tau];\\
  0, & t>2T+\tau.
\end{array}
\right.
\end{equation}
The above RZ pulses are characterized by following parameters: $v_0$
is the maximum strength of the coupling, $T$ is the scanning period
of RZ pulse, and $\tau$ is an alterable time interval between two
pulses.

This scheme is analogous  to a normal Ramsey interferometer while
the Ramsey pulses at the beginning and the end of the sequence that
couple the two components and  redistribute the populations on each
component are replaced  by so called nonlinear RZ tunneling process
\cite{Ye}. The two tunneling processes are separated by a holding
period. During the holding period, there is no coupling between the
two components and the BEC on each component will evolve
independently and only acquire different additional phases. In the
course of the simulative experiments, the system is prepared in one
internal state initially, the final populations of atoms in each
state are recorded when the second pulse turns off. The measurements
are repeated with variable time interval $\tau$. The final
populations are sensitive to the phase difference built up between
two components during the intermediate
 period, as a result,  the Ramsey fringes pattern is expected to emerge in
 time domain.

\subsection{Ramsey fringe patterns}

The nonlinear Schr\"{o}dinger equations (\ref{seq}) that govern the
temporal evolution of the two-component BEC system are solved
numerically using standard Runge-Kutta 4-5th algorithm. We set the
initial condition $(a,b)=(1,0)$, and take the maximum coupling
strength
  as the energy scale, namely,
$v_0=1$. The Ramsey fringe patterns have been obtained by recording
the final transition probability $|b|^2$ versus the holding time
$\tau$.

\begin{figure}[h]
\begin{center}
\rotatebox{0}{\resizebox *{8.0cm}{6.cm} {\includegraphics
{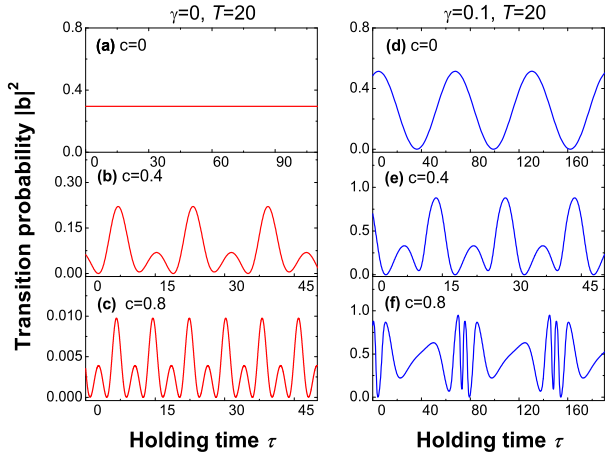}}}
\end{center}
\caption{(Color online) Ramsey fringe patterns for symmetric case
(left column) and asymmetric case (right column) under different
nonlinear parameters with $T=20$. (a) and (d) $c=0$, (b) and (e)
$c=0.4$, (c) and (f) $c=0.8$.}\label{fringec}
\end{figure}

We begin our numerical simulations with the linear case of $c=0$ for
$T=20$. Figs. \ref{fringec}(a) and \ref{fringec}(d) shows the
variation of the transition  probability for symmetric case
($\gamma=0$) and asymmetric case ($\gamma=0.1$), respectively.
Actually Eq. (\ref{seq}) can be solved analytically for the
symmetric case, the solution is $\sin^2(v_0T/2)$ which depends on
the scanning period $T$ only. The numerical result in Fig.
\ref{fringec}(a) coincides with the analytic prediction that the
transition probability keeps a constant $0.29596$. For the
asymmetric system the standard Ramsey fringes pattern of typical
sinusoidal is shown as Fig. \ref{fringec}(d).

\begin{figure}[h]
\begin{center}
\rotatebox{0}{\resizebox *{8.0cm}{6.cm} {\includegraphics
{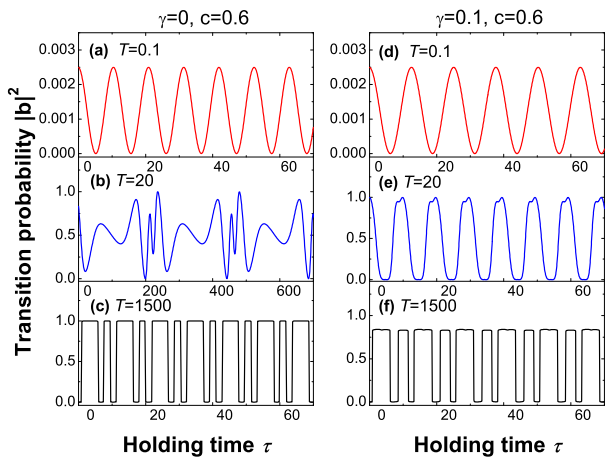}}}
\end{center}
\caption{(Color online) Ramsey fringe patterns for symmetric case
(left column) and asymmetric case (right column) under different
scanning periods with $c=0.6$. (a) and (d) $T=0.1$, (b) and (e)
$T=20$, (c) and (f) $T=1500$.}\label{fringet}
\end{figure}

With the emergence of nonlinearity, the Ramsey fringes pattern
distinctly deviates from that of linear case due to the dramatic
changes of the transition dynamics. In this case the system
(\ref{seq}) is no longer analytically solvable. Our numerical
simulations for different nonlinear parameters and  various scanning
periods of the RZ pulse have been displayed in Figs. \ref{fringec}
and \ref{fringet}, respectively. Fig. \ref{fringec} show that both
nonlinearity and symmetry can affect the pattern and the frequency
of Ramsey fringes significantly. By analyzing the results in Figs.
\ref{fringec} and \ref{fringet} we find that the Ramsey fringes
pattern includes perfect sinusoidal or cosinoidal oscillation [see
Figs. \ref{fringec}(d), \ref{fringet}(a) and \ref{fringet}(d)],
trigonometric oscillation with multiple period [see Figs.
\ref{fringec}(b), \ref{fringec}(c), \ref{fringec}(e),
\ref{fringec}(f), \ref{fringet}(b), and \ref{fringet}(e)], and
rectangular oscillation [see Figs. \ref{fringet}(c) and
\ref{fringet}(f)]. Furthermore, we also find that the sinusoidal
Ramsey pattern only exists in the linear case ($c=0$) and the rapid
scanning case ($T=0.1$) while the rectangular oscillation only
emerges in the very slow scanning case ($T=1500$).  These
diversiform interference patterns are distinguished  from the normal
Ramsey fringes of sinusoidal or cosinoidal forms and are obviously
evoked by the nonlinear atomic interaction.

\section{ Theoretical Analysis and Extended Numerical Simulation}
In this section we will present thorough analysis on these striking
interference patterns. In practical experiments, in contrast to the
oscillating amplitudes and shapes of the fringe patterns, the
frequencies of the patterns are of more interest and could be
recorded with relatively high resolution and contrast, therefore we
focus our theoretical analysis on the frequency property extracted
from the Ramsey interference patterns through  the Fourier
transformation (FT). We find that the frequencies of patterns that
are dramatically modulated by the interplay of nonlinearity and
symmetry and contain many information about the intrinsic properties
of the BEC system.

Through investigating the nonlinear Ramsey patterns presented above
we see the time scale of the period of the RZ pulse plays an
important role in forming the striking patterns. So our following
discussions are divided into two limit cases, i.e., sudden limit and
adiabatic limit.  In the former case, the time scale of the RZ pulse
is fast compared to the intrinsic motion of the system that is
characterized by the frequency $v_0$, while the adiabatic limit
refers to the case that the RZ pulse is much slower than intrinsic
motion of the system.

\subsection{Sudden limit case, i.e., $T\ll2\pi/v_0$. }

\begin{figure}[h]
\begin{center}
\rotatebox{0}{\resizebox *{7.5cm}{12.cm} {\includegraphics
{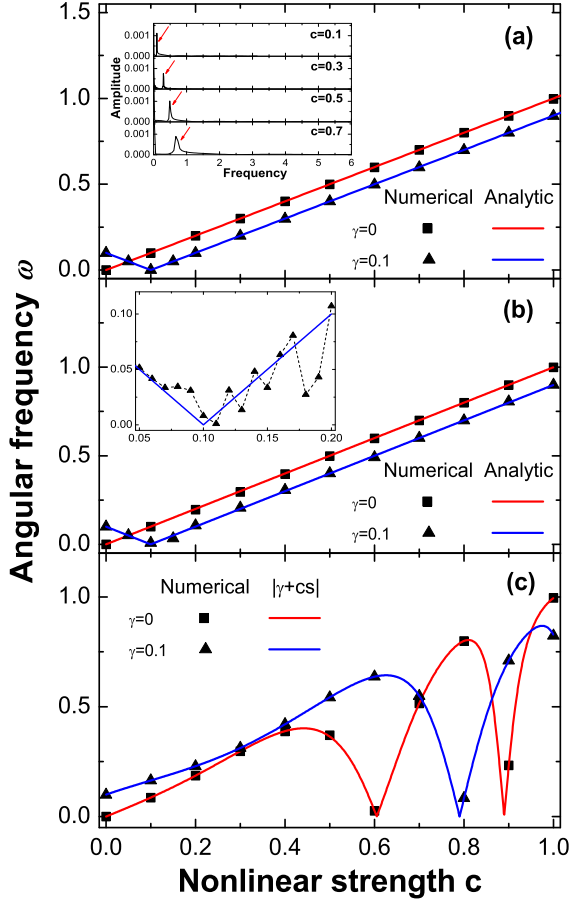}}}
\end{center}
\caption{(Color online) Angular frequency $\omega$ of Ramsey fringes
as a function of the nonlinear strength $c$. The numerical results
well agree with analytic predictions. (a) sudden limit case, the
inset displays the frequency spectrum of Ramsey fringes obtained
from Fourier transformation for different $c$ with $\gamma=0$ (red
arrows refer to the numerical results plotted in main plot). (b)
adiabatic limit case, the inset demonstrates the details for
$\gamma=0.1$ with $c$ from 0.05 to 0.2. (c) general
situation.}\label{allwc}
\end{figure}

In our simulation, we choose  the scanning period of the RZ pulse
$T$ as $0.1$ that is much smaller than the intrinsic period of the
system $2\pi/v_0$ .  For both symmetric and asymmetric cases we
extract the angular frequency information of the Ramsey fringes
through making the  FT analysis on the data. The results have been
demonstrated in Fig. \ref{allwc}(a). A perfect linear increase
relation between the angular frequencies of Ramsey fringes and
nonlinear parameters is shown for symmetric case [see the solid
squares in Fig. \ref{allwc}(a)]. For the asymmetric case, the
frequency decreases linearly and then increases linearly as the
nonlinear strength increases [see the solid triangles in Fig.
\ref{allwc}(a)]. The dip to zero at $c=\gamma=0.1$ is clearly seen
in the asymmetric system.

Now we explain the above numerical results through some analytic
deduction. Considering that the transition probability from one
state to the other state is small enough in the sudden limit, thus
we can use the perturbation method to analyze the system
(\ref{seq}). We introduce the following variable transformation
\begin{eqnarray}
&&a=a'\exp\left[-i\int_{0}^{t}\left(\frac{\gamma}{2}+\frac{c}{2}(|
b|^{2}-|a|^{2})\right)dt\right],\\ \label{transform1}
&&b=b'\exp\left[i\int_{0}^{t}\left(\frac{\gamma}{2}+\frac{c}{2}(|
b|^{2}-|a|^{2})\right)dt\right]. \label{transform2}
\end{eqnarray}%
Following this transformation, we  transform the diagonal terms in
the Hamiltonian (\ref{H}) away and obtain the first-order amplitude
of $b(T)$ which yields $b(T)=\int_{0}^{t}\frac{v_{0}}{2}\sin
^{2}(\frac{\pi t}{T})\mathrm{e}^{i(c-\gamma)t}dt|_{t=T}$. Finally,
the transition probability after the first RZ pulse is given by
\begin{equation}
|b(T)|^2=\frac{2\pi ^{4}v_{0}^{2}[1-\cos (\Omega
T)]}{\Omega^{2}(4\pi ^{2}-\Omega^{2}T^{2})^{2}},
\label{suddenformula}
\end{equation}%
where $\Omega=c-\gamma$. For convenience, we introduce a phase shift
$\phi(\tau)$ to describe the different phase accumulations between
two components during the holding period. Considering that two
components evolve independently during this period, we get
$\phi(\tau)=|\gamma+cs|\tau$ from Eq. (\ref{seq}), where
$s=|b(T)|^2-|a(T)|^2$ denotes the population difference between two
components when the first pulse has been turned off. This phase
shift is proportional to the holding time. Obviously, the angular
frequency of the Ramsey fringes is expected to  be
\begin{equation}
\omega=|\gamma+cs|.
\end{equation}
 This result implies that the
frequency of Ramsey fringes is entirely determined by the population
difference $s$ and the parameters $\gamma$ and $c$. Substituting Eq.
(\ref{suddenformula}) into the above formula, we obtain the angular
frequency of Ramsey fringes in the form
\begin{equation}\label{sudomega}
\omega=\left\vert\frac{4cv_0^2\pi^4[1-\cos(\Omega
T)]}{\Omega^2[4\pi^2-(\Omega T)^2]^2}-\Omega \right\vert.
\end{equation}
The above analytical predictions are compared with our numerical
results in Fig. \ref{allwc}(a) and a perfect agreement is shown.
Indeed, under the sudden limit assumption, the term $\Omega T$ in
Eq. (\ref{sudomega}) is a small quantity, the numerator of the first
term on the right-hand side of Eq. (\ref{sudomega}) is close to zero
due to $\cos(\Omega T)\rightarrow 1$. When $T\rightarrow0$, one can
safely neglect the first term on the right-hand of Eq.
(\ref{sudomega}), then the frequency is proportional to the
parameter $|\Omega|$.


\subsection{Adiabatic limit case, i.e., $T\gg2\pi/v_0$.}

In order to ensure the scanning period long enough, we set $T$ as
$1500$ in calculation. In contrast to the linear case and the sudden
limit case, an important phenomenon in this case is found that the
FT on Ramsey fringes reveals multiple frequency components, namely,
$\omega=n\omega_0$, where $\omega_0$ is the fundamental frequency
(i.e., basic or first frequency) of the fringes, $n$ is a positive
integer. We interpret this in terms of the interplay between
nonlinearity ascribed to the interatomic interaction and the
coupling energy from the external laser field.
 Fig. \ref{allwc}(b) only illustrates the fundamental frequencies
 of
Ramsey fringes for different nonlinear parameters.

\begin{figure}[h]
\begin{center}
\rotatebox{0}{\resizebox *{7.5cm}{5.50cm} {\includegraphics
{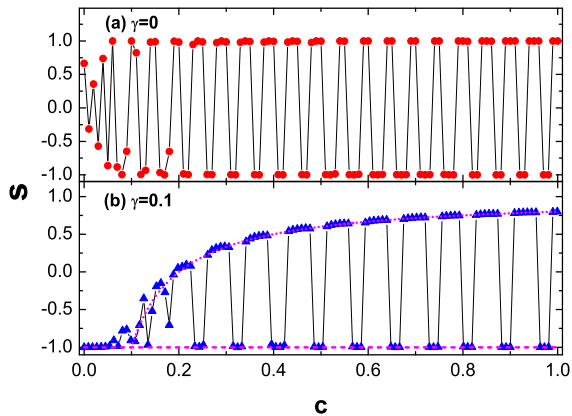}}}
\end{center}
\caption {(Color online) The population difference $s$ versus
nonlinear
 parameters from 0 to 1 for symmetric case (red circles)
 and asymmetric case (blue triangles) with $T=1500$. The dotted and dashed lines refer to
 theoretical prediction from Eq. (\ref{sdt1}).}\label{adsc}
\end{figure}

The results in this case are very similar to that in sudden limit
case. However, a novel phenomenon is that there is a irregular
fluctuation in near $c=\gamma$ region [see the inset in Fig.
\ref{allwc}(b)]. We guess the adiabatic assumption is violated in
this region. To confirm this argument, we trace the population
difference $s$ after the first RZ pulse with nonlinear parameter
increasing. The results are presented in Fig. \ref{adsc}, we see
that a irregular oscillation of $s$ occurs in the region where
$|\gamma-c|$ is very small as well. With the nonlinear parameter
increasing from 0.25 to 1,
 $s$ will jump between two points $+1$ and $-1$ in symmetric case.
However, for asymmetric system, when $c>0.35$, the value of $s$ will
jump between $-1$ and another unknown point. This is a more
intriguing quantum phenomenon and more essentially physical reasons
 need further detailed study.
\begin{figure}[h]
\begin{center}
\rotatebox{0}{\resizebox *{7.5cm}{5.50cm} {\includegraphics
{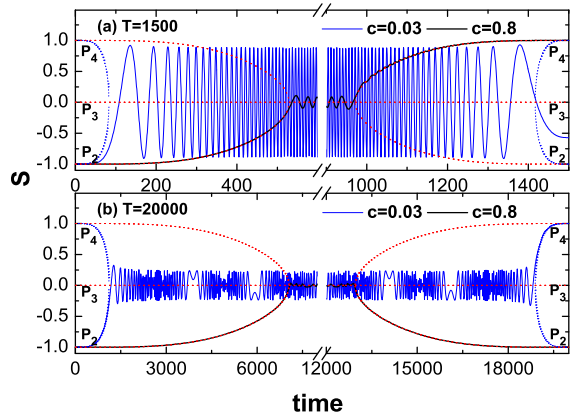}}}
\end{center}
\caption
 {(Color online) Comparison between the dynamical evolution (solid line)
 and the adiabatic evolution (dotted line) of fixed points for
 symmetric case with different $T$: (a) $1500$, (b) $20000$. Blue line and black line
refer to $c=0.03$ and $c=0.8$, respectively. Blue dotted line and
red dotted line show
  the corresponding adiabatic evolution obtained from Eq. (\ref{fixed}).}\label{trace1}
\end{figure}

In order to explain the above peculiar phenomena, under the
mean-field approximation, following Ref. \cite{clasicalH}, we
introduce the relative phase $\theta=\theta_b-\theta_a$ and the
population difference $s=|b|^2-|a|^2$ as two canonical conjugate
variables, then we can obtain an effective classical Hamiltonian
\begin{equation}\label{CH}
 \mathcal{H}=-(\gamma+{c\over
2}s)s+v\sqrt{1-s^2}\cos\theta.
\end{equation}
This classical Hamiltonian can describe completely the dynamic
properties of system (\ref{seq}) \cite{clasicalH}. The adiabatic
evolution of the quantum eigenstates can be evaluated by tracing the
shift of the classical fixed points in phase space when the
parameter $v$ varies in time slowly \cite{fixedpoint}. According to
Refs. \cite{Ye,prefu}, for symmetric system we get the classical
fixed points on line $\theta^\ast=\pi$,
\begin{equation}
s^\ast=\left\{
\begin{array}{ll}
  0, & c/v<1; \\
  0,\pm\sqrt{1-(v/c)^2}, & c/v>1.
\end{array}\label{fixed}
\right.
\end{equation}

We show the evolution of fixed point $s^\ast=-1$ ($P_2$) in Fig.
\ref{trace1}. The three fixed points in Eq. (\ref{fixed}) are
characterized by $P_3$, $P_4$ and $P_2$, respectively. One saddle
point $P_3$ ($s^\ast=0$) and two elliptic points $P_2$ and $P_4$
correspond to one unstable state and two stable states. For $c=0.8$,
a good agreement between dynamical evolution and adiabatic
trajectory of $P_2$ is shown both for $T=1500$ and $T=20000$.
However, for $c=0.03$, the evolution of fixed point $P_2$ shows a
clear deviation from the adiabatic trajectory given by Eq.
(\ref{fixed}) at $T=1500$ [see Fig. \ref{trace1}(a)] while the fixed
point can follow the adiabatic evolution at $T=20000$ [see Fig.
\ref{trace1}(b)]. The phenomena indicate that the adiabatic
condition cannot be satisfied for $c=0.03$ where occurs the
irregular fluctuation at $T=1500$ in Fig. \ref{adsc}. Therefore, we
give the adiabatic condition as follows:
\begin{eqnarray}\label{adiabatic}
T\gg\mathrm{Max}\left[\frac{2\pi}{|\gamma-c|},\frac{2\pi}{v_0}\right].
\end{eqnarray}

\begin{figure}[h]
\begin{center}
\rotatebox{0}{\resizebox *{7.5cm}{5.50cm} {\includegraphics
{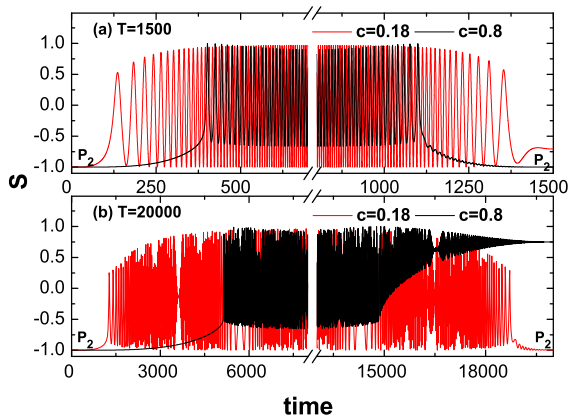}}}
\end{center}
\caption
 {(Color online) Evolution of fixed points for
 asymmetric case under different $T$: (a) $1500$, (b) $20000$.}\label{trace2}
\end{figure}

Under this condition, so long as $\gamma\neq c$, the system will
evolve adiabatically if the scanning period is long enough even for
the small nonlinear parameters \cite{prefu}. This can successfully
explain the novel fluctuation in Figs. \ref{allwc}(b) and
\ref{adsc}. Accordingly, we trace the fixed point $P_2$ in
asymmetric case (see Fig. \ref{trace2}) using same parameter $T$ as
in Fig. \ref{trace1}. The similar feature that good adiabatic
evolution for $c=0.8$ and nonadiabatic evolution for $c=0.18$ where
is in the close vicinity of the zero-energy resonance ($\gamma=c$)
with $T=1500$ is observed. In addition, another interesting
phenomenon is also find, despite the evolution process of the fixed
point is not clear, there are two final states of adiabatic
evolution to be choose for the fixed point [see Fig. \ref{trace2}(b)
and Fig. \ref{adsc}(b)] for asymmetric case. We will interpret it by
some deeply physical analysis below.

 For the adiabatic limit case, the energy of system both for symmetric and asymmetric cases
 is no longer conservative during the entire
evolution process, however at the beginning and end of the evolution
the corresponding energies of the system keep the same value,
\begin{equation}\label{H0T}
\mathcal{H}(s=-1,t=0)=\mathcal{H}(s^\ast,t=T).
\end{equation}
In our scheme, both for $t=0$ and $t=T$, the coupling parameter
$v=0$. Thus we can get the final state of system from Eqs.
(\ref{CH}) and (\ref{H0T})
\begin{equation}\label{sdt1}
s^\ast=\left\{
\begin{array}{ll}
                  -1, & \gamma>c;\\
                  -1, 1-2\gamma/c, & 0<\gamma<c.
                \end{array}
\right.
\end{equation}

This result implies that,at the end of the adiabatic evolution, the
system has two states to choose when $c>\gamma$ for this case, one
choice is back to the initial state $s^\ast=-1$ and the other choice
is located on another state of the identical energy with the initial
state $s^\ast=1-2\gamma/c$. However the latter choice restricts the
population to $|b|^2=\gamma/c$, in other words, the quantum
tunneling for asymmetric case require the atom number on another
state must be not more than $N\gamma/c$ ($N$ is the total number of
atoms). We use the above analysis to check our numerical results in
Fig. \ref{adsc}(b) and a good agreement is shown. According to this
analytic prediction, in adiabatic limit case, the final value of $s$
should be $-0.11$ or $-1$ for $c=0.18$ and $0.75$ or $-1$ for
$c=0.8$ in Fig. \ref{trace2}, these results strongly support our
numerical results.
\begin{figure}[h]
\begin{center}
\rotatebox{0}{\resizebox *{8.5cm}{5.cm} {\includegraphics
{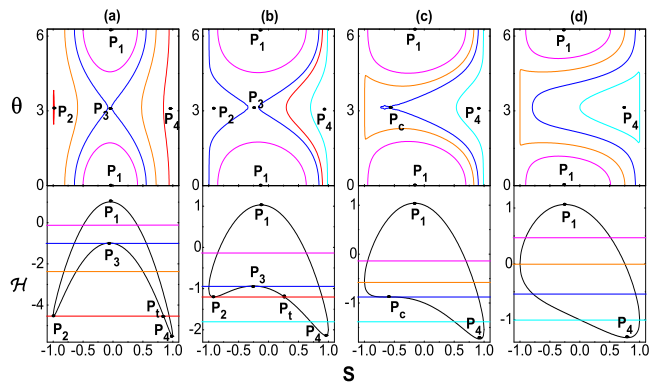}}}
\end{center}
\caption
 {(Color online) Evolution of the phase space motions as $c/v$ changes adiabatically (upper panel) with $\gamma=0.5$. (a)
 $c/v=10$, (b) $c/v=3$, (c) $c/v=2.0897$, (d) $c/v=1$. The lower panel is the corresponding energy curve
 for
 $\theta=0$ (black thin line) and $\pi$ (black heavy line).}\label{Phase}
\end{figure}

In order to provide a simple intuitive understand of this adiabatic
evolution process, we study the evolution of fixed points in phase
space as shown in Fig. \ref{Phase}. $P_1$, $P_2$ and $P_4$ in the
upper panel of Fig. \ref{Phase} are all elliptic points
corresponding to the local maximum ($P_1$) and minimum ($P_2$ and
$P_4$) of the classical Hamiltonian indicated in the lower panel of
Fig. \ref{Phase}, respectively. We see the quantum transition
 between two states can be explained by a collision between
two fixed points. When $c/v$ decreases from $10$ to $1$, the fixed
point $P_2$ will collide with the unstable saddle point $P_3$ at
$P_c$ and disappear subsequently, as shown in Figs.
\ref{Phase}$(a)\rightarrow(b)\rightarrow(c)\rightarrow(d)$. The
condition of the collision is given by Ref. \cite{fixedpoint},
namely,
\begin{equation}
v=(c^{2/3}-\gamma^{2/3})^{3/2}.
\end{equation}
For the case with $\gamma=0.5$, the collision occurs at $c/v=2.0897$
[see Fig. \ref{Phase}(c)]. However, when $c/v$ increases from 1 to
10 again, the state of system will choose either stable fixed point
$P_2$ or a stable trajectory $P_t$ which is of identical energy with
$P_2$ to follow after the dynamical bifurcation at $P_c$ [see Figs.
\ref{Phase}$(c)\rightarrow(b)\rightarrow(a)$]. This is a peculiar
and intriguing phenomenon that only emerges in asymmetric system.
Following the above analysis, we can obtain the analytic expression
of fundamental frequency of Ramsey fringes in adiabatic limit from
Eqs. (\ref{sudomega}) and (\ref{sdt1})
\begin{equation}
\omega=|\gamma-c|.
\end{equation}
The results show a perfect linear relation both for symmetric and
asymmetric cases and are consistent with our numerical results [see
\ref{allwc}(b)].



\subsection{General situation}

In this subsection, we turn to study the general case where the
scanning period of RZ pulse is of the same order with $2\pi/v_0$,
i.e., $T=20$. We will show the population difference $s$ can greatly
affect the frequency of Ramsey fringes in this case. Similarly, we
show the fundamental frequencies of Ramsey fringes in Fig.
\ref{allwc}(c). The comparison
  between numerical results and theoretical prediction $|\gamma+cs|$ show
  a good agreement. In Fig.
 \ref{allwc}(c), the perfect linear relation has been completely broken,
and three zero-frequency points emerge: one
  in asymmetric case and two in symmetric case. The physics behind
 this is that the balance between energy difference characterized
  by $\gamma$ and the interatomic interaction energy
  controlled by the nonlinear term $cs$.
When the nonlinear parameters satisfy
  the balance condition $\gamma=-cs$, there will occur zero-energy resonance
  or the zero-frequency points.

  To confirm this argument, we trace the
  population difference with the nonlinear parameter increasing.
The results show that, for symmetric case when two components are of
identical populations, the Ramsey fringes vanish and the
zero-frequency points emerge. The concrete process of evolution of
system in general case is not clear due to the complex quantum
transition behaviors.


\subsection{The dependence of frequency of $\gamma$}

In this part, we briefly investigate the case which sets the
nonlinear parameter as a constant and takes $\gamma$ as an alterable
quantity.
\begin{figure}[h]
\begin{center}
\rotatebox{0}{\resizebox *{7.5cm}{12.0cm} {\includegraphics
{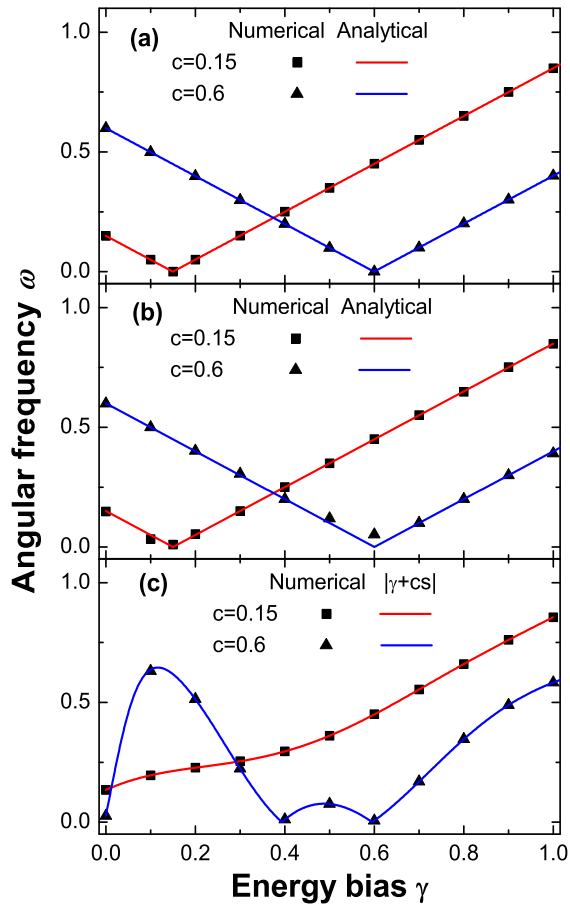}}}
\end{center}
\caption
 {(Color online) The angular frequency of Ramsey fringes versus
  the energy difference $\gamma$ for different cases.
 (a) sudden limit; (b) adiabatic limit;
 (c) general situation.}\label{gamma}
\end{figure}

Following the previous analysis, the fundamental frequency of Ramsey
fringes is also expected to be $\omega=|\gamma+cs|$. Fig.
\ref{gamma} shows the fundamental frequencies of Ramsey fringes
versus energy difference $\gamma$ for different scanning periods. We
have used the same parameter $T$ as in Fig. \ref{allwc}, and Figs.
\ref{gamma}(a), (b), and (c) refer to the sudden limit, the
adiabatic limit and the general case, respectively.

By analyzing these plots, we see that, there is a common property
for three cases, zero-frequency points emerge when the nonlinear
parameter equals to the energy difference for large nonlinear
parameter $c=0.6$. However, for small nonlinear parameter $c=0.15$,
there does not occur zero-frequency points in general case while
zero-frequency points emerge in sudden limit and adiabatic limit
cases. Here, we restrict our consideration to $\gamma>0$ and $c>0$.
In fact, we find the zero-frequency point in general case occurs at
$\gamma=-0.118$ for $c=0.15$, and the zero-frequency point in
general case is more than one.

In particular, the similar irregular fluctuation in the region
around $\gamma=c$ has been found in Fig. \ref{gamma}(b). The smaller
the nonlinear parameter is, the larger the amplitude of irregular
oscillation shows. This implies that in the region around
$\gamma/c=1$, the system does not satisfy the adiabatic condition
(\ref{adiabatic}). If the scanning period is long enough, the novel
fluctuation in Fig. \ref{gamma}(b) will become smooth \cite{prefu}.


\section{Discussions and Applications}

In summary, based on the quantum Rosen-Zener tunneling process, we
propose a feasible scheme to realize nonlinear Ramsey interferometry
with a two-component Bose-Einstein condensate, where the
nonlinearity arises from the interaction between coherent atoms. In
our scheme, two RZ pulses are separated by an intermediate holding
period of variable duration and through varying the holding period
we have observed nice Ramsey fringe patterns in time domain. In
contrast to the standard Ramsey fringes our nonlinear Ramsey
patterns display diversiform structures due to the interplay of the
nonlinearity and asymmetry. In particular, we find that the
frequency of the nonlinear Ramsey fringes exactly reflects the
strength of nonlinearity as well as the asymmetry of system. Our
study suggests that our interferometry scheme can be used to measure
the atomic parameters such as scattering length, atom number and
energy spectrum  through measuring the frequency of nonlinear Ramsey
interference fringe patterns.

Our nonlinear Ramsey interferometer scheme can also be realized
using the BECs with a double-well potential. This BEC system, under
the mean-field approximation, is described by following
Gross-Pitaevskii equation (GPE)
\begin{equation}
i\hbar\frac{\partial\Psi(r,t)}{\partial
t}=\left[-\frac{\hbar^2}{2m}\nabla^2+V(r)+U_0|\Psi(r,t)|^2\right]\Psi(r,t),
\end{equation}
where $U_0= 4\pi\hbar^2a_sN/m$ with $m$ the atomic mass and $a_s$
 the $s-$wave scattering length of the atoms. The wave
 function can be described by a superposition of two states
that localize in each well separately as \cite{smerzi}
$\Psi(r,t)=\psi_1(t)\phi_1(r)+\psi_2(t)\phi_2(r).$ The spatial wave
function $\phi_i(r)$ ($i=1,2$) which describe the condensate in each
well can be expressed in terms of symmetric and antisymmetric
 stationary eigenstates of GPE, and these two wave functions satisfy
 the orthogonality condition $\int\phi_1(r)\phi_2(r)dr=0$ and normalized condition
 $\int|\phi_i(r)|^2dr=1$. Consider the weakly linked BEC, the dynamic behavior of system
  can be described by Schr\"{o}dinger equation with the Hamiltonian as follows:
\begin{equation}
H= \left(
\begin{array}{cc}
      \epsilon^0_1+c_1|\psi_1|^2 & K \\
      K & \epsilon^0_2+c_2|\psi_2|^2 \\
\end{array}
\right),
\end{equation}
where
$\epsilon^0_i=\int[\frac{\hbar^2}{2m}|\nabla\phi_i|^2+|\phi_i|^2V(r)]dr$
($i=1,2$) is the zero-point energy in each well.
$\Delta\epsilon=\epsilon_1-\epsilon_2$ is the energy bias.
$c_i=U_0\int|\phi_i|^4dr$ denotes the atomic self-interaction.
$K=\int[\frac{\hbar^2}{2m}(\nabla\phi_1\nabla\phi_2)+\phi_1V(r)\phi_2]dr$
stands for the the amplitude of the coupling between two wells.

For example, consider one dimension case, we can express the
potential of our system as $V({x})=\frac{1}{2}m\omega
x^2+v_0\mathrm{e}^{-{x^2}/{2d}}+fx$, $d$ is the double-well
separation in $x$ direction. This optical double-well potential can
be created by superimposing a blue-detuned laser beam upon the
center of the magnetic trap \cite{2well}, the difference of the
zero-point energy between two wells or trap asymmetry characterized
by $f$ can be bringed by a magnetic field, a gravity field or light
shifts \cite{gravity}. The atomic interaction $c$ can be adjusted
flexibly by Feshbach resonance, and the barrier height $K$ can be
effectively controlled by adjusting the intensity of the
blue-detuned laser beam.

\bigskip

\section{Acknowledgments}
This work is supported by National Natural Science Foundation of
China (No.10725521,10604009,10875098), the National Fundamental
Research Programme of China under Grant No. 2006CB921400,
2007CB814800.

\end{document}